\documentclass[12pt,aps,prb,preprint,eqsecnum]{revtex4}

\usepackage{graphics}

\begin{document}
\title{Bracket relations for relativity groups}
\author{Thomas F. Jordan}
\email[email: ]{tjordan@d.umn.edu}
\affiliation{Physics Department, University of Minnesota, Duluth, Minnesota 55812}
 
\begin{abstract}
Poisson bracket relations for generators of canonical transformations are derived directly from the Galilei and Poincar\'e groups of changes of space-time coordinates. The method is simple but rigorous. The meaning of each step is clear because it corresponds to an operation in the group of changes of space-time coordinates. Only products and inverses are used; differences are not used. It is made explicitly clear why constants occur in some bracket relations but not in others, and how some constants can be removed, so that in the end there is a constant in the bracket relations for the Galilei group but not for the Poincar\'e group. Each change of coordinates needs to be only to first order, so matrices are not needed for rotations or Lorentz transformations; simple three-vector descriptions are enough.

Conversion to quantum mechanics is immediate. One result is a simpler derivation of the commutation relations for angular momentum directly from rotations. Problems are included.

\end{abstract}


\maketitle

\section{Introduction}\label{one}

The bracket relations of the Galilei group and the Poincar\'e group are expressions of relativity in the working language of Hamiltonian dynamics. The relativity is that of Galileo and Newton for the Galilei group in ``nonrelativistic'' mechanics and that of Einstein for the Poincar\'e group in ``relativistic'' mechanics. Our goal here is to understand the foundation of these bracket relations in classical Hamiltonian mechanics when the brackets are Poisson brackets and the generators in the bracket relations are functions of the canonical coordinates and momenta that generate canonical transformations. We will see how to get from the groups of changes of space-time coordinates to the Poisson-bracket relations for the generators of canonical transformations.

The derivation of the bracket relations for the rotation generators may be the most used part. Almost every course in quantum mechanics uses the commutation relations for angular momentum and says something about their connection to rotations. That connection is described very directly and simply here, particularly in Section V.B and the paragraph that contains Eqs.~(\ref{jjnoc}) and (\ref{jjnoc2}). 
It is clear that quantum mechanics does not play an essential role. The equations can be read equally well as quantum mechanics or classical mechanics. It is clear that properties of rotations give the bracket relations only to within constants. The constants can be removed from the bracket relations only by making choices of the constants that can be added to the generators.

Extension to the Galilei group or Poincar\'e group brings more applications. When the bracket relations of the generators with the position are included, to ensure that the way the generators change the position corresponds to the way the relativity group changes coordinates, major elements of classical and quantum mechanics can be derived. The ``nonrelativistic'' and ``relativistic'' forms of the Hamiltonian can be found for an object in classical mechanics.\cite{me74} These give the relation of the canonical momentum to the velocity and the interpretation of the canonical momentum and Hamiltonian as physical quantities. We can see how the translation generator is related to mass and velocity and why $-i\nabla $ represents the momentum in quantum mechanics.\cite{me29} We can show that an interaction potential for two particles can depend only on the relative position and momentum, not on the center-of-mass position and momentum.\cite{meLinearOperators} Some of these results are reviewed in Section VII. They are all consequences of relativistic symmetries. They are obtained from the bracket relations. The bracket relations provide a structure that underlies and shapes both classical mechanics and quantum mechanics and is simple enough that it does not depend on either. Our derivation of the bracket relations here can be used in teaching both classical and quantum mechanics.

The method used here is simple but rigorous. The meaning of each step is clear because it corresponds to an operation in the group of changes of space-time coordinates. Only products and inverses are used; differences are not used. A framework that supports sums and differences comes in when we think of functions of the canonical coordinates and momenta as physical quantities and obtain their brackets from the brackets of the canonical coordinates and momenta. That is completely separate from what we are doing here. We have two different ways to get the same bracket equations. For example, we can get the bracket relations for angular momentum by looking at angular-momentum functions as generators of rotations, as we are doing here, or by looking at them as physical quantities made from positions and momenta and using the bracket relations for position and momentum. Our understanding and appreciation will be aided, and our teaching will be clearer, when each method is presented without mixing in operations from the other.

It is made explicitly clear why constants occur in some bracket relations but not in others, and how some constants can be removed, so that in the end there is a constant in the bracket relations for the Galilei group but not for the Poincar\'e group. Problem 7.5 is stated to show that this constant makes nonrelativistic mechanics unable to describe conservation of momentum without conservation of mass in radioactive decays that were observed before Einstein presented his relativity.

Conversion to quantum mechanics is simple for most of what is done here. The equations remain the same for finding the bracket relations with possible constants in Section V and for eliminating constants in Section VI. What we see when we read the equations is changed, and we use different language to describe it. For the generators and the quantities being transformed, we see Hermitian operators instead of real functions of the canonical coordinates and momenta. The brackets are commutators divided by $i$ instead of Poisson brackets. There are unitary transformations instead of canonical transformations. The constants that can occur in bracket relations and be added to generators are multiples of the identity operator. The converted procedure is in the Heisenberg picture of quantum mechanics.

I wrote this in response to students in my class in classical mechanics asking for ``something to read about this.'' Something clear and simple was needed. Derivations of bracket relations in classical mechanics are found in more advanced treatments.\cite{SudarshanMukunda,PauriProsperi75} Derivations in quantum mechanics are in the Schr\"odinger picture. They cannot be simply converted to a familiar form of classical mechanics. The 
Schr\"odinger picture also brings in the ambiguity about phase factors of state vectors. To set up the closest Schr\"odinger-picture analog\cite{meLinearOperators} of what is done here, it is necessary to at least observe that the phase ambiguity in the product of two changes of state does not depend on the state vector, and prove that the phase factors can be eliminated for the one-parameter subgroups. An alternative for the Poincar\'e group is to prove that all the phase factors that could affect bracket relations can be eliminated.\cite{halpern68} For the Galilei group, that can be done only in an extension to a larger group.\cite{Bargmann54,Voisin65} 

The basic one-parameter subgroups of space and time translations, rotations, and Galilei or Lorentz transformations are the most familiar parts of the Galilei and Poincar\'e groups. The bracket relations of their generators show how these ten one-parameter subgroups fit together in the larger groups. For the Galilei or Poincar\'e group, a bracket of two of the ten generators of the basic one-parameter subgroups never contains a linear combination of more than one of these generators; if it is not zero or a constant, it is just plus or minus one of the ten generators. Derivations of bracket relations in quantum mechanics most often use products of three transformations, with the last the inverse of the first and the middle one infinitesimal. Then the product rule shows how the generators are transformed, and that shows what the bracket relations are. This can be done beautifully,\cite{weinberg95} but seeing what comes from the product rule requires identification of  generators for more than the ten basic one-parameter subgroups. Here we use products of four transformations, with the third the inverse of the first and the fourth the inverse of the second.\cite{meLinearOperators,vanderWaerden74} This pulls out the bracket directly and gives just the one of the ten generators, or the minus one or none, that is the answer. There is no intermediate step that requires identification of another generator.  

The procedure here is made simpler than the closest quantum parallel \cite{meLinearOperators} by understanding that, although results are obtained in second order, each step needs to be carried only to first order. This allows simple three-vector calculations. In particular, in Section V.B, matrices are not needed for rotations or rotation generators. These simplifications can be made in quantum mechanics as well as in classical mechanics.

\section{Changes of coordinates}\label{two}

Our description of a physical system uses a time coordinate $t$, and at each time $t$ the position of each object in space is described by a coordinate vector $\vec r$ with components \mbox{$x$, $y$, $z$} along axes in three orthogonal directions. We consider several different changes of coordinates:
\vspace{0.6cm}

\noindent\textit{Rotations.} The frame of orthogonal axes for the space coordinates is rotated around an axis through the origin. The position of each object is described by a new coordinate vector $\vec r \, '$ with components $x'$, $y'$, $z'$ along the new axes.
\vspace{0.6cm}

\noindent\textit{Space translations.} The origin for the space coordinates is moved a fixed distance $-\vec d$. Each space coordinate vector is changed from $\vec r$ to
\begin{equation}
\label{st}
\vec r \, ' =\vec r + \vec d.
\end{equation}

\noindent\textit{Galilei transformations.} The position of each object is described by a new coordinate vector
\begin{equation}
\label{Gal}
\vec r \, ' =\vec r - \vec \beta t
\end{equation}
relative to an origin moving with velocity $\vec \beta $.
\vspace{0.6cm}

\noindent\textit{Time translations.} The time coordinate is changed from t to
\begin{equation}
\label{timetrans}
t' = t - s
\end{equation}
as it would be for clocks set ahead by a fixed amount $s$.
\vspace{0.6cm}

We consider all the changes of coordinates that can be made from these rotations, space translations, Galilei transformations, and time translations. They form a group.
The group product of two changes of coordinates is the change of coordinates obtained by doing first one and then the other. This is called the \textit{Galilei group}.
\newpage

For example, a time translation
\begin{equation}
\label{timetrans2}
\vec r \, ' =\vec r , \quad \quad t' = t - s
\end{equation}
followed by a Galilei transformation
\begin{equation}
\label{Gal2}
\vec r \, '' = \vec r \, ' - \vec \beta t', \quad \quad t'' = t'
\end{equation}
gives
\begin{equation}
\label{timetrans2Gal2prod}
\vec r \, '' =\vec r - \vec \beta t + \vec \beta s, \quad \quad t'' = t -s.
\end{equation}
The product of a time translation and a Galilei transformation includes a space translation; the distance $\vec d$ of the space translation is $\vec \beta s$.

We also consider the \textit{Poincar\'e group}. It is obtained by replacing the Galilei transformations with Lorentz transformations. For example, a Galilei transformation with $\vec \beta $ in the $z$ direction is replaced by the Lorentz transformation
\begin{eqnarray}
\label{Lorentztrans}
x' & = & x, \quad \quad z' = z\cosh \alpha  - t \sinh \alpha  \nonumber \\
y' & = & y, \quad \quad t' = t\cosh \alpha  - z \sinh \alpha
\end{eqnarray}
where $|\vec \beta |= \tanh \alpha $ is the velocity $dz/dt$ of the origin of the $x'$, $y'$, $z'$ coordinates. We use units where the velocity of light $c$ is $1$.

We will look at these groups as composites of one-parameter subgroups. A one-parameter subgroup is a set of changes of coordinates that depend on a parameter $u$ so that when $u$ is zero there is no change of coordinates, the identity element of the group, and the product of two changes of coordinates for values $u_1$ and $u_2$ of the parameter is the change of coordinates for $u_1+u_2$. The time translations form a one-parameter subgroup for which the parameter is $s$. Rotations around a fixed axis form a one-parameter subgroup for which the parameter is the angle of rotation. Space translations in a fixed direction form a one-parameter subgroup for which the parameter is the distance (positive or negative). Galilei transformations for velocities in a fixed direction form a one-parameter subgroup for which the parameter is the velocity. Lorentz transformations for velocities in a fixed direction form a one-parameter subgroup. For velocities in the $z$ direction, for example, they are described by Eqs.~(\ref{Lorentztrans}). The parameter is $\alpha $; this will be shown in Problem 2.3. The Galilei group, or the Poincar\'e group, is a composite of ten one-parameter subgroups: time translations, rotations around the $x$, $y$, and $z$ axes, space translations in the $x$, $y$, and $z$ directions, and Galilei transformations or Lorentz transformations for velocities in the $x$, $y$, and $z$ directions. The way these ten one-parameter subgroups fit together gives the bracket relations for the ten generators $H$, $\vec P$, $\vec J$, and $\vec G$ or $\vec K$ to be introduced in Section IV.
\vspace{0.6cm}

\noindent\textbf{Problem 2.1.} Every change of coordinates in the Galilei group can be written as
\begin{equation}
\label{Galgeneral}
\vec r \, ' = R\vec r  + \vec d  - \vec \beta t, \quad \quad t' = t - s
\end{equation}
where $R$ denotes a rotation of the vector $\vec r$. Show this by showing that the product of two changes of coordinates of this form is a change of coordinates of the same form: if the first is for $R_1$, $\vec d_1$, $\vec \beta _1$, $s_1$ and the second is for $R_2$, $\vec d_2$, $\vec \beta _2$, $s_2$, then the product is for  $R_2R_1$, $R_2\vec d_1+\vec \beta _2s_1+\vec d_2$, $R_2\vec \beta _1+\vec \beta _2$, $s_1+s_2$ with $R_2R_1$ the product of the rotations $R_1$ and $R_2$. Equations (\ref{timetrans2Gal2prod}) provide an example.
\vspace{0.6cm}

\noindent\textbf{Problem 2.2.} Show that the inverse of the change of coordinates described by Eqs.~(\ref{Galgeneral}) in Problem 2.1 is the change of coordinates described by the same equations for $R^{-1}$, \mbox{$R^{-1}\vec \beta s-R^{-1}\vec d$}, $-R^{-1}\vec \beta $, $-s$ with $R^{-1}$ the inverse of the rotation $R$.
\vspace{0.6cm}

\noindent\textbf{Problem 2.3.} Show that the Lorentz transformations described by Eqs.~(\ref{Lorentztrans}) form a one-parameter group for which $\alpha $ is the parameter.

\section{Relativity}\label{three}

We assume that each of these changes of coordinates leads to an equivalent description of the system and its dynamics; physical quantities and the way they change can be described with the new coordinates as well as with the old. In quantum mechanics, we assume that for each change of coordinates there is a unitary operator that changes the operators that represent physical quantities in the Heisenberg picture, and also changes the Hamiltonian operator that generates the changes described by dynamics. In classical Hamiltonian mechanics, we assume that for each change of coordinates there is a canonical transformation that changes the canonical coordinates and momenta, and also changes the Hamiltonian function that generates the changes described by dynamics. 

When the change of coordinates is a time translation, the canonical transformation is the same as for a change in time described by the dynamics. The change of canonical coordinates and momenta from $q_n$, $p_n$ to $q'_n$, $p'_n$ corresponding to the change of the time coordinate from $t$ to $t'$ is simply that
\begin{equation}
\label{canfort}
q'_n(t') = q_n(t), \quad \quad p'_n(t') = p_n(t).
\end{equation}
There is no change beyond referral to a different time. When $t'$ is $t-s$, this means that
\begin{equation}
\label{canfort2}
q'_n(t') = q_n(t'+ s), \quad \quad p'_n(t') = p_n(t'+s).
\end{equation}

We assume that after one canonical transformation for one change of coordinates, a second canonical transformation can be made the same way for a second change of coordinates. This means that the product of the two canonical transformations, the result of following one by the other, is the canonical transformation that corresponds to the product of the two changes of coordinates. This assumption is not always valid. For example, if there is a time-dependent external force, the canonical transformations that make the changes in time described by the dynamics will be different from one time to another. The assumption holds at least for closed systems that are isolated from outside influences.

\section{Generators}\label{four}

We consider classical Hamiltonian dynamics for a  Hamiltonian $H$ that is a function of the canonical coordinates and momenta and does not depend on time. A function $F$ of the canonical coordinates and momenta that represents a physical quantity at time zero is changed by the dynamics between time zero and time $t$ to a function $F(t)$ of the canonical coordinates and momenta that is determined by the equation of motion \cite{me75,DasFerbel}
\begin{equation}
\label{eqnmotion}
\frac{dF(t)}{dt} = [F(t), H]
\end{equation}
and the boundary condition that $F(t)$ is $F$ at time zero. We write $[F,G]$ for the Poisson bracket of any two functions $F$ and $G$ of the canonical coordinates and momenta. \cite{me75,DasFerbel} When the series converges, the solution of the equation of motion (\ref{eqnmotion}) is
\begin{equation}
\label{seriessolution}
F(t) = F + t[F,H] + \frac{1}{2}t^2[[F,H],H]... + \frac{1}{k!}t^k[...[F,H]...H] +...
\end{equation}
in which the bracket with $H$ is taken $k$ times in the term with $t^k$. The change of functions of the canonical coordinates and momenta between time zero and time $s$ is a canonical transformation.\cite{me75} The same canonical transformation makes the change determined by the dynamics between time $t'$ and $t'+s$. This is the canonical transformation that corresponds to the time translation, the change in the time coordinate from $t$ to $t-s$, as is explained in the discussion leading to  Eqs.~(\ref{canfort2}). As a function of $s$, these canonical transformations form a one-parameter group; this is to be shown in Problem 4.1. 

The Hamiltonian function $H$ generates a one-parameter group of canonical transformations. Conversely, every one-parameter group of canonical transformations has a generator, a function of the canonical coordinates and momenta, that acts like the Hamiltonian; this is shown in the Appendix. For each of the ten basic one-parameter groups of changes of coordinates, we assume there is an identified one-parameter group of corresponding canonical transformations. We give the generators different names. The Hamiltonian $H$ is the generator of the canonical transformations that correspond to time translations. We let $P_1$, $P_2$, $P_3$ be the generators of the one-parameter groups of canonical transformations that correspond to space translations in the $x$, $y$, $z$ directions, let $J_1$, $J_2$, $J_3$ be the generators of the one-parameter groups of canonical transformations that correspond to rotations around the $x$, $y$, $z$ axes, let $G_1$, $G_2$, $G_3$ be the generators of the one-parameter groups of canonical transformations that correspond to Galilei transformations for velocities in the $x$, $y$, $z$ directions, and let $K_1$, $K_2$, $K_3$ be the generators of the one-parameter groups of canonical transformations that correspond to Lorentz transformations for velocities in the $x$, $y$, $z$ directions. We write $\vec P$, $\vec J$, $\vec G$, and $\vec K$ for the sets of three generators. We are making plus and minus sign conventions by saying that the canonical transformations generated by $H$, $\vec P$, $\vec G$, and $\vec K$ with signs as in Eq.~(\ref{seriessolution}) correspond to the changes of coordinates with the signs in Eqs.~(\ref{timetrans}), (\ref{st}), (\ref{Gal}), and (\ref{Lorentztrans}). The sign convention for $\vec J$ is that the canonical transformations generated by $J_1$, for example, correspond to
\begin{eqnarray}
\label{rot}
y' & = & y\cos \theta  - z\sin \theta   \nonumber \\
z' & = & z\cos \theta  + y\sin \theta .
\end{eqnarray}

A one-parameter group of canonical transformations does not completely determine a generator; adding a constant to a generator does not change the transformations it generates. The transformations do determine a generator to within addition of constants; adding a function of the canonical coordinates and momenta to a generator does change the transformations it generates when the added function is not a constant.

The one-parameter group of canonical transformations that corresponds to a one-parameter group of changes of coordinates can be identified by seeing that the way the canonical transformations change particular functions of the canonical coordinates and momenta that describe particular physical quantities is the way those quantities are supposed to be changed by the changes of coordinates. Examples are worked out in Section VII.
\vspace{0.6cm}

\noindent\textbf{Problem 4.1.} For each function $F$ of the canonical coordinates and momenta let $C_t(F)$ be the $F(t)$ given by Eq.~(\ref{seriessolution}). Show that
\begin{equation}
\label{1pgroup}
C_u(C_t(F)) = C_{t+u}(F).
\end{equation}
Hint: Compare the power series you get for $C_u(C_t(F))$ with the
\begin{center}
(Power Series)$\times $(Power Series) = Power Series 
\end{center}
for $e^te^u = e^{t+u}$. This shows that the canonical transformations generated by a Hamiltonian form a one-parameter group.

\section{Bracket relations}\label{five} 

We will show that, when the correct constants are added to them, the generators $H$, $\vec P$, $\vec J$, $\vec G$ for the Galilei group satisfy the Poisson-bracket relations
\begin{equation}
\label{bracrel}
[J_j, J_k] = \epsilon_{jkm}J_m, \quad \quad  [J_j, P_k] = \epsilon_{jkm}P_m,
\end{equation}
\begin{equation}
\label{Gbracrel2} 
[J_j, G_k] = \epsilon_{jkm}G_m, \quad \quad  [G_j, H] = P_j,
\end{equation}  
\begin{equation}
\label{Gbracrel3}
[G_j, P_k] = \delta_{jk}M
\end{equation}
with $M$ a real number, the generators $H$, $\vec P$, $\vec J$, $\vec K$ for the Poincar\'e group satisfy the Poisson-bracket relations (\ref{bracrel}), the same as for the Galilei group, and 
\begin{equation}
\label{Kbracrel2} 
[J_j, K_k] = \epsilon_{jkm}K_m, \quad \quad  [K_j, H] = P_j,
\end{equation}  
\begin{equation}
\label{Kbracrel3}
[K_j, K_k] = -\epsilon_{jkm}J_m, \quad \quad  [K_j, P_k] = \delta_{jk}H,
\end{equation}
and all the other Poisson brackets of the generators $H$, $\vec P$, $\vec J$, and $\vec G$ or $\vec K$ are zero. First we show that each bracket relation is true if a constant is added to its right side. Then, in Section VI, we will show that all the constants except $M$ can be eliminated when constants are added to the generators.

\subsection{The pattern}\label{A}

Our first calculation of a Poisson bracket of two generators will be a pattern for the others. To find $[G_1, H]$, we consider a product of four infinitesimal canonical transformations made from generators $G_1$, $H$, $G_1$, $H$ with parameters $\epsilon $, $\delta $, $-\epsilon $, $-\delta $. We calculate power series to second order in $\epsilon $ and $\delta $. If either $\epsilon $ or $\delta $ is zero, the product of the four canonical transformations will be just the product of one canonical transformation and its inverse, which is the identity transformation, the same as when $\epsilon $ and $\delta $ are both zero, so the lowest-order terms in the power series for the product will be proportional to $\epsilon \delta $, not $\epsilon $, $\epsilon^2$, $\delta $, or $\delta^2$. We need to go only to first order in $\epsilon $ and first order in $\delta $. The product of the four canonical transformations takes each function $F$ of the canonical coordinates and momenta through the sequence of transformations
\begin{eqnarray}
\label{canprod}
 F & \rightarrow  & F' \, \, \, = F + \epsilon [F, G_1]     \nonumber \\
   & \rightarrow  & F'' \, \, = F' + \delta [F', H']     \nonumber \\
   & \rightarrow  & F''' \, = F'' - \epsilon [F'', G''_1]     \nonumber \\
   & \rightarrow  & F'''' = F''' - \delta [F''', H'''].
\end{eqnarray}
The generators are changed the same as any other functions of the canonical coordinates and momenta. To the first order that we need, the transformed generators are
\begin{eqnarray}
\label{gentrans}
     H' & = & H + \epsilon [H, G_1]     \nonumber \\
  G''_1 & = & G_1 + \delta [G_1, H]     \nonumber \\
   H''' & = & H.
\end{eqnarray}
Using these, the Jacobi identity
\begin{equation}
\label{Jacobi}
[[A,B],C] = [[C,B],A] + [[A,C],B],
\end{equation}  
and the antisymmetry property of the Poisson bracket,
\begin{equation} 
\label{antisym}
[A,B] = -[B,A],
\end{equation} 
we find that the result of the product of the four canonical transformations described by Eq.~(\ref{canprod}) is that
\begin{equation}
\label{canprodr}
F \rightarrow  F'''' = F - \epsilon \delta [F, [G_1. H]].
\end{equation}
This is the same as the lowest-order term of the canonical transformation generated by $[G_1, H]$ with parameter $-\epsilon \delta $. We assume this must be the lowest-order term of the canonical transformation that corresponds to the product of the four changes of coordinates that correspond to the four canonical transformations generated by $G_1$ and $H$. By calculating this product of changes of coordinates, we can find $[G_1, H]$. 

The sequence of four changes of coordinates corresponding to the four canonical transformations generated by $G_1$, $H$, $G_1$, and  $H$ with parameters $\epsilon $, $\delta $, $-\epsilon $, $-\delta $ the same as for the canonical transformations in Eq.~(\ref{canprod}) is
\begin{eqnarray}
\label{GHxtprod}
x, t  & \rightarrow  & x' \, \, \, = x - \epsilon t, \quad \, \, \, \,   t' = t    \nonumber \\
      & \rightarrow  & x'' \, \, = x', \quad \quad \quad \, t'' = t' - \delta     \nonumber \\
      & \rightarrow  & x''' \,  = x'' + \epsilon t'', \, \,  t'''  = t''    \nonumber \\
      & \rightarrow  & x'''' = x''', \quad \quad \, \,  t'''' = t''' + \delta 
\end{eqnarray}
with no changes in $y$ and $z$. The product of these four changes of coordinates, which is the result of the sequence, is
\begin{equation}
\label{GHxtresult}
x \rightarrow  x'''' = x - \epsilon \delta 
\end{equation}
with no change in $y$, $z$ and $t$. It is a space translation in the $x$ direction. The canonical transformation that corresponds to this change of coordinates gives
\begin{equation}
\label{sptr}
F \rightarrow  F - \epsilon \delta [F, P_1]
\end{equation}
to lowest order. Comparing with Eq.~(\ref{canprodr}), we see that $[G_1, H]$ and $P_1$ must generate the same canonical transformations. This implies that $[G_1, H]$ is either $P_1$ or $P_1$ plus a constant. We can show similarly that $[G_j, H]$ is either $P_j$ or $P_j$ plus a constant for $j=1,2,3$. A similar calculation for $[K_j, H]$ is to be done as Problem 5.3. 

To find the Poisson bracket of another pair of generators, we can use Eq.~(\ref{canprodr}) with $G_1$ and $H$ replaced by that pair, and calculate the product of the corresponding four changes of coordinates. When the changes of coordinates commute, the product of the four changes of coordinates is the identity change of coordinates, no change at all, and the generator of the corresponding canonical transformation is either zero or a constant. Thus we find that each of the Poisson brackets
\begin{equation}
\label{0brac}
[H, P_k], \quad \quad  [H, J_k], \quad \quad [P_j, P_k],
\end{equation}
\begin{equation}
\label{0brac2}
[P_k, J_k], \quad \quad  [G_k, J_k], \quad \quad  [K_k, J_k],
\end{equation}
\begin{equation}
\label{0brac3}
[G_j, G_k], \quad \quad  [G_j, P_k],
\end{equation}
is either zero or a constant. For all of these brackets except the last, the different changes of coordinates commute simply because they change different coordinates.

When a calculation is needed, it needs to be only to second order in $\epsilon $ and $\delta $. We need to go only to first order in $\epsilon $ and first order in $\delta $; just as in the product of the four canonical transformations, if either $\epsilon $ or $\delta $ is zero, the product of the four changes of coordinates will be just the product of one change and its inverse, which is the identity change, the same as when $\epsilon $ and $\delta $ are both zero, so the lowest-order terms in the power series for the product will be proportional to $\epsilon \delta $, not $\epsilon $, $\epsilon^2$, $\delta $, or $\delta^2$.

\subsection{Rotations}\label{B}

To first order, the rotation described by Eq.~(\ref{rot}) is also described by
\begin{equation}
\label{vecrot}
\vec r \, ' = \vec r  + \epsilon \, \hat x\hspace{-0.1cm} \times \hspace{-0.1cm} \vec r
\end{equation}
with $\epsilon $ the infinitesimal value of the angle $\theta $. We write $\hat x $, $\hat y $, $\hat z $ for unit vectors in the $x$, $y$, $z$ directions. To find the Poisson bracket $[J_1, J_2]$, we calculate the product of four rotations, around the $x$, $y$, $x$, $y$ axes, through the angles $\epsilon $, $\delta $, $-\epsilon $, $-\delta $:
\begin{eqnarray}
\label{rotprod}
 \vec r & \rightarrow  & \vec r \, ' \, \, \, = \vec r + \epsilon \, \hat x\hspace{-0.1cm} \times \hspace{-0.1cm} \vec r     \nonumber \\
   & \rightarrow  & \vec r \, '' \, \, = \vec r \, ' + \delta \, \hat y\hspace{-0.1cm} \times \hspace{-0.1cm} \vec r \, '     \nonumber \\
   & \rightarrow  & \vec r \, ''' \, = \vec r \, '' - \epsilon \, \hat x\hspace{-0.1cm} \times \hspace{-0.1cm} \vec r \, ''     \nonumber \\
   & \rightarrow  & \vec r \, '''' = \vec r \, ''' - \delta \, \hat y\hspace{-0.1cm} \times \hspace{-0.1cm} \vec r \, '''.
\end{eqnarray}
To lowest order, the result is 
\begin{eqnarray}
\label{rotprod2}
 \vec r \rightarrow \vec r \, '''' & = & \vec r - \epsilon \delta \, \hat x\hspace{-0.1cm} \times \hspace{-0.1cm} (\hat y\hspace{-0.1cm} \times \hspace{-0.1cm} \vec r)  + \epsilon \delta \, \hat y\hspace{-0.1cm} \times \hspace{-0.1cm} (\hat x\hspace{-0.1cm} \times \hspace{-0.1cm} \vec r)   \nonumber \\
   & = & \vec r - \epsilon \delta \, (\hat x\hspace{-0.06cm}\cdot \hspace{-0.06cm}\vec r) \, \hat y  + \epsilon \delta \, (\hat y\hspace{-0.06cm}\cdot \hspace{-0.06cm}\vec r) \, \hat x     \nonumber \\
   & = & \vec r - \epsilon \delta \, (\hat x\hspace{-0.1cm} \times \hspace{-0.1cm} \hat y)\hspace{-0.1cm} \times \hspace{-0.1cm} \vec r     \nonumber \\
   & = & \vec r - \epsilon \delta \, \hat z \hspace{-0.1cm} \times \hspace{-0.1cm} \vec r.
\end{eqnarray}
The change of coordinates that is the product of the four rotations is rotation by $-\epsilon \delta $ around the $z$ axis. The corresponding canonical transformation gives
\begin{equation}
\label{jjtr}
F \rightarrow  F - \epsilon \delta [F, J_3]
\end{equation}
in the lowest order. Comparing this with Eq.~(\ref{canprodr}) with $G_1$ and $H$ replaced by $J_1$ and $J_2$, we conclude that $[J_1, J_2]$ is either $J_3$ or $J_3$ plus a constant. We can find similarly that the Eq.~(\ref{bracrel}) for each $[J_j, J_k]$ is true with a constant added.

\subsection{Rotations and translations}\label{C}

For $[J_1, P_2]$, the product of the four changes of coordinates is
\begin{eqnarray}
\label{jpprod}
 \vec r & \rightarrow  & \vec r \, ' \, \, \, = \vec r + \epsilon \, \hat x\hspace{-0.1cm} \times \hspace{-0.1cm} \vec r     \nonumber \\
   & \rightarrow  & \vec r \, '' \, \, = \vec r \, ' + \delta \, \hat y     \nonumber \\
   & \rightarrow  & \vec r \, ''' \, = \vec r \, '' - \epsilon \, \hat x\hspace{-0.1cm} \times \hspace{-0.1cm} \vec {r''}     \nonumber \\
   & \rightarrow  & \vec r \, '''' = \vec r \, ''' - \delta \, \hat y.
\end{eqnarray}
To lowest order, the result is that
\begin{eqnarray}
\label{jpprod2}
 \vec r \rightarrow \vec r \, '''' & = & \vec r - \epsilon \delta \, (\hat x\hspace{-0.1cm} \times \hspace{-0.1cm} \hat y)     \nonumber \\
   & = & \vec r - \epsilon \delta \, \hat z.
\end{eqnarray}
It is a space translation in the $z$ direction. The corresponding canonical transformation gives
\begin{equation}
\label{jptr}
F \rightarrow  F - \epsilon \delta [F, P_3]
\end{equation}
to lowest order. Comparing this with Eq.~(\ref{canprodr}) for $J_1$ and $P_2$, we conclude that $[J_1, P_2]$ is either $P_3$ or $P_3$ plus a constant. We can do a similar calculation for each $[J_j, P_k]$ with $j$ and $k$ different. For $[J_k, P_k]$, the changes of coordinates commute, because the rotation and the space translation change different coordinates, so the product of the four changes of coordinates is no change at all,  and the generator of the corresponding canonical transformation is either zero or a constant. Thus we find that the Eq.~(\ref{bracrel}) for each $[J_j, P_k]$ is true with a constant added. Similar calculations for $[J_j, G_k]$ and $[J_j, K_k]$ are to be done as Problems 5.1 and 5.2. 

\subsection{Lorentz transformations}\label{D} 

To first order, the Lorentz transformation described by Eq.~(\ref{Lorentztrans}) is also described by
\begin{equation}
\label{vecLtrans}
\vec r \, ' = \vec r  - \epsilon \, \hat z \, t, \quad \quad 
t' = t - \epsilon \, \hat z\hspace{-0.06cm}\cdot \hspace{-0.06cm}\vec r
\end{equation}
with $\epsilon $ the infinitesimal value of the parameter $\alpha $. For $[K_1, K_2]$, the product of the four changes of coordinates is
\begin{eqnarray}
\label{kkprod}
 \vec r, \, t & \rightarrow  & \vec r \, ' \, \, \, = \vec r - \epsilon \, \hat x \, t, \quad \, \, \, \, \, \, \, \, \, \, \, t' = t - \epsilon \, \hat x\hspace{-0.06cm}\cdot \hspace{-0.06cm}\vec r  
   \nonumber \\
   & \rightarrow  & \vec r \, '' \, \, = \vec r \, ' - \delta \, \hat y \, t',  \, \, \, \, \, \, \quad t'' = t' - \delta \, \hat y\hspace{-0.06cm}\cdot \hspace{-0.06cm}\vec r \, '  \nonumber \\
   & \rightarrow  & \vec r \, ''' \, = \vec r \, '' + \epsilon \, \hat x \, t'',  \, \, \, \quad  t''' = t'' + \epsilon \, \hat x\hspace{-0.06cm}\cdot \hspace{-0.06cm}\vec r \, '' \nonumber \\
   & \rightarrow  & \vec r \, '''' = \vec r \, ''' + \delta \, \hat y \,  t''', \quad t'''' = t''' + \delta \, \hat y\hspace{-0.06cm}\cdot \hspace{-0.06cm}\vec r \, '''.
\end{eqnarray}
To lowest order, the result is that
\begin{eqnarray}
\label{kkprod2}
 \vec r \rightarrow \vec r \, '''' & = & \vec r - \epsilon \delta \, (\hat y\hspace{-0.06cm}\cdot \hspace{-0.06cm}\vec r) \, \hat x  + \epsilon \delta \, (\hat x\hspace{-0.06cm}\cdot \hspace{-0.06cm}\vec r) \, \hat y     \nonumber \\
   & = & \vec r - \epsilon \delta \, (\hat y\hspace{-0.1cm} \times \hspace{-0.1cm} \hat x)\hspace{-0.1cm} \times \hspace{-0.1cm} \vec r     \nonumber \\
   & = & \vec r + \epsilon \delta \, \hat z \hspace{-0.1cm} \times \hspace{-0.1cm} \vec r
\end{eqnarray}
and $t$ is not changed. The product of the four Lorentz transformations is rotation by $\epsilon \delta $ around the $z$ axis. The corresponding canonical transformation gives
\begin{equation}
\label{jptr}
F \rightarrow  F + \epsilon \delta [F, J_3]
\end{equation}
to lowest order. Comparing this with Eq.~(\ref{canprodr}) for $K_1$ and $K_2$, we conclude that $[K_1, K_2]$ is either $-J_3$ or $-J_3$ plus a constant. We can find similarly that the Eq.~(\ref{Kbracrel3}) for each $[K_j, K_k]$ is true with a constant added.

\subsection{Lorentz transformations and translations}\label{E}

For $[K_3, P_3]$, the product of the four changes of coordinates is
\begin{eqnarray}
\label{kpprod}
 z, \, t & \rightarrow  & z' \, \, \, = z - \epsilon \, t, \quad \, \, \, \, \, \, t' = t - \epsilon \, z  
   \nonumber \\
   & \rightarrow  & z'' \, \, = z' + \delta ,  \, \, \, \, \, \, \quad t'' = t'  \nonumber \\
   & \rightarrow  & z''' \, = z'' + \epsilon \, t'',  \, \, \, \, t''' = t'' + \epsilon \, z'' \nonumber \\
   & \rightarrow  & z'''' = z''' - \delta , \quad \, t'''' = t'''
\end{eqnarray}
with no changes in $x$ and $y$. To lowest order, the result is that $x$, $y$, $z$ are not changed and
\begin{equation}
\label{kpprod2}
 t \rightarrow t'''' = t + \epsilon \delta .
\end{equation}
It is a time translation. The corresponding canonical transformation gives
\begin{equation}
\label{kptr}
F \rightarrow  F - \epsilon \delta [F, H]
\end{equation}
to lowest order. Comparing this with Eq.~(\ref{canprodr}) for $K_3$ and $P_3$, we conclude that $[K_3, P_3]$ is either $H$ or $H$ plus a constant. We can find similarly that the Eq.~(\ref{Kbracrel3}) for each $[K_k, P_k]$ is true with a constant added. For $[K_j, P_k]$ with $j$ and $k$ different, the four changes of coordinates commute, so their product is no change at all, and the generator of the corresponding canonical transformation is either zero or a constant. Thus we find that the Eq.~(\ref{Kbracrel3}) for each $[K_j, P_k]$ is true with a constant added.

At every step, we see a another new simplification that the method brings. Because each change of coordinates needs to be only to first order, matrices are not needed for either rotations or Lorentz transformations. It is enough to use simple three-vector descriptions, and they fit readily with translations. When the calculations to be done in problems are included, we can see that every one of the bracket relations for the Galilei group and the Poincar\'e group is true with a constant added.
\vspace{0.6cm}

\noindent\textbf{Problem 5.1.} Use language from Eqs.~(\ref{Gal}) and (\ref{jpprod}) to calculate the product of the four changes of coordinates for $[J_1, G_2]$ and see that $[J_1, G_2]$ is either $G_3$ or $G_3$ plus a constant. Consider the similar calculation for $[J_1, G_1]$ and see that $[J_1, G_1]$ is either zero or a constant. We can find similarly that the Eq.~(\ref{Gbracrel2}) for each $[J_j, G_k]$ is true with a constant added.
\vspace{0.6cm}

\noindent\textbf{Problem 5.2.} Use language from Eqs.~(\ref{jpprod}) and (\ref{kkprod}) for the product of the four changes of coordinates to show that $[J_1, K_2]$ is either $K_3$ or $K_3$ plus a constant. Show that $[J_1, K_1]$ is zero or a constant and that the Eq.~(\ref{Kbracrel2}) for each $[J_j, K_k]$ is true with a constant added.
\vspace{0.6cm}

\noindent\textbf{Problem 5.3.} Calculate the product of the four changes of coordinates for $[K_1, H]$ and see that $[K_1, H]$ is either $P_1$ or $P_1$ plus a constant. This can be done simply by putting the Lorentz transformations of time coordinates from Eqs.~(\ref{kkprod}) into Eqs.~(\ref{GHxtprod}); this is all that is needed to change the Galilei transformations to Lorentz transformations. We can find similarly that the Eq.~(\ref{Kbracrel2}) for each $[K_j, H]$ is true with a constant added.

\section{Eliminating constants}\label{six}

We have shown that every one of the bracket relations for the Galilei group and the Poincar\'e group is true with a constant added. Now we will show that all the constants except $M$ can be eliminated when constants are added to the generators.

The presence of constants is limited because the Poisson brackets are antisymmetric and satisfy the Jacobi identity. For example, from the antisymmetry (\ref{antisym}), the Jacobi identity (\ref{Jacobi}), and bracket relations possibly with constants, we can see that
\begin{eqnarray}
\label{ppnoc}
[P_3, P_1] & = & [[J_1, P_2], P_1]     \nonumber \\
           & = & [[P_1, P_2], J_1] + [[J_1, P_1], P_2]     \nonumber \\
           & = & 0
\end{eqnarray}
because inside the brackets the constants give the same result as zero. We can calculate similarly that
\begin{equation}
\label{ggnoc}
     [P_j, P_k] = 0,  \quad [G_j, G_k] = 0,
\end{equation}
\begin{equation}
\label{gphnoc}
      \, \,  [P_j, H] = 0,  \quad \, \, \, [J_j, H] = 0.
\end{equation}

From the antisymmetry of the brackets, we have
\begin{equation}
\label{jjnoc}
     [J_j, J_k] = \epsilon_{jkm}J_m + \epsilon_{jkm}b_m
\end{equation}
where $b_1$, $b_2$, $b_3$ are real numbers. By adding these constants to the generators $J_1$, $J_2$, $J_3$, we get
\begin{equation}
\label{jjnoc2}
     [J_j, J_k] = \epsilon_{jkm}J_m.
\end{equation}

From the antisymmetry, the Jacobi identity, and bracket relations possibly with constants, we get
\begin{eqnarray}
\label{jjpjnoc}
[J_3, P_3] & = & [[J_1, J_2], P_3]     \nonumber \\
           & = & [[P_3, J_2], J_1] + [[J_1, P_3], J_2]     \nonumber \\
           & = & [J_1, P_1] + [J_2, P_2]
\end{eqnarray}
and similarly
\begin{equation}
\label{jjpjnoc2}
[J_1, P_1] = [J_2, P_2] + [J_3, P_3]  
\end{equation}
from which we see that $[J_2, P_2]$ is zero and conclude that similarly each $[J_k, P_k]$ is zero. In the same way, we get
\begin{eqnarray}
\label{jjpjnoc3}
[J_2, P_3] & = & [[J_3, J_1], P_3]     \nonumber \\
           & = & [[P_3, J_1], J_3] + [[J_3, P_3], J_1]     \nonumber \\
           & = & -[J_3, P_2]
\end{eqnarray}
and conclude that each $[J_j, P_k]$ is $-[J_k, P_j]$ so that
\begin{equation}
\label{jjpjnoc4}
     [J_j, P_k] = \epsilon_{jkm}P_m + \epsilon_{jkm}b_m
\end{equation}
with real numbers $b_1$, $b_2$, $b_3$. By adding these constants to the generators $P_1$, $P_2$, $P_3$, we get
\begin{equation}
\label{jjpjnoc5}
     [J_j, P_k] = \epsilon_{jkm}P_m.
\end{equation}
We can see similarly that by adding constants to the generators $G_1$, $G_2$, $G_3$ and  $K_1$, $K_2$, $K_3$ we can get 
\begin{equation}
\label{jjgjnoc5}
     [J_j, G_k] = \epsilon_{jkm}G_m
\end{equation}
and
\begin{equation}
\label{jjkknoc5}
     [J_j, K_k] = \epsilon_{jkm}K_m.
\end{equation}

In the same way, we get
\begin{eqnarray}
\label{ghnoc}
[G_3, H] & = & [[J_1, G_2], H]     \nonumber \\
           & = & [[H, G_2], J_1] + [[J_1, H], G_2]     \nonumber \\
           & = & [J_1, P_2] = P_3
\end{eqnarray}
and conclude that, similarly,
\begin{equation}
\label{ghnoc2}
     [G_j, H] = P_j
\end{equation}
and
\begin{equation}
\label{khnoc}
     [K_j, H] = P_j.
\end{equation}

In the same way, we get
\begin{eqnarray}
\label{gpnoc}
[G_3, P_1] & = & [[J_1, G_2], P_1]     \nonumber \\
           & = & [[P_1, G_2], J_1] + [[J_1, P_1], G_2]     \nonumber \\
           & = & 0
\end{eqnarray}
and
\begin{eqnarray}
\label{gpnoc2}
[G_3, P_3] & = & [[J_1, G_2], P_3]     \nonumber \\
           & = & [[P_3, G_2], J_1] + [[J_1, P_3], G_2]     \nonumber \\
           & = & [G_2, P_2]
\end{eqnarray}
and conclude that
\begin{equation}
\label{gpnoc3}
     [G_j, P_k] = \delta_{jk}M
\end{equation}
with $M$ a real number. The same equations (\ref{gpnoc}) and (\ref{gpnoc2}) hold with $\vec{K}$ in place of $\vec{G}$. From them, we conclude that 
\begin{equation}
\label{kpnoc}
     [K_j, P_k] = \delta_{jk}H  + \delta_{jk}M
\end{equation}
with $M$ a real number. By adding $M$ to $H$ we get
\begin{equation}
\label{kpnoc2}
     [K_j, P_k] = \delta_{jk}H.
\end{equation}

The one remaining step, to see that there are no constants in the equations for $[K_j, K_k]$, is to be done as Problem 6.1. When that is included, we can see that all the constants except the $M$ for $[G_k, P_k]$ can be eliminated by adding constants to the generators. When that is done, the generators $\vec P$, $\vec J$, $\vec G$, and $\vec K$ are completely determined; we cannot add constants to them without putting constants back in the bracket relations. For the Poincar\'e group, $H$ is also completely determined; we can not add a constant to $H$ without putting a constant back in the equation for $[K_k, P_k]$. For the Galilei group, $H$ is not completely determined. We can still add a constant to $H$. It will not change the bracket relations for the Galilei group because in them $H$ never occurs outside on the right. In examples, we will see that the $M$ of $[G_k, P_k]$ is a mass; it is the nonrelativistic limit of the $H$ of $[K_k, P_k]$.

Removing the constants puts the bracket relations into the simple standard forms that are generally used. When we look at examples in Section VII, we will see that the adjustment of constants leaves the generators in familiar forms that can be identified with physical quantities. Then the bracket relations for the generators correspond to bracket relations for physical quantities. Constants in the bracket relations for generators do not change the group structure, but constants in the bracket relations for physical quantities can be important, as we know from the example of Planck's constant in the commutation relations for position and momentum.
\vspace{0.6cm}

\noindent\textbf{Problem 6.1.} Use Eq.~(\ref{jjnoc2}) for $[J_2, J_1]$, other bracket relations possibly with constants, the antisymmetry of Poisson brackets, and the Jacobi identity applied to $[[K_1, K_3], J_1]$, to show that $[K_1, K_2]$ is $-J_3$ without a constant. We can conclude that similarly $[K_j, K_k]$ is $-\epsilon_{jkm}J_m$ without a constant.

\section{Examples}\label{seven}

Consider a single object moving in three-dimensional space. There are three degrees of freedom. We use canonical coordinates and momenta $q_1$, $q_2$, $q_3$ and $p_1$, $p_2$, $p_3$ that are components of three-dimensional vectors $\vec{q}$ and $\vec{p}$.  Functions of the canonical coordinates and momenta that satisfy the bracket relations are
\begin{eqnarray}
\label{ggenfunc}
     \vec{P} & = & \vec{p}, \quad \quad \, \, \vec{J} = \vec{q}\hspace{-0.1cm} \times \hspace{-0.1cm}\vec{p},    \nonumber \\
  \vec{G} & = & M\vec{q}, \quad H = \frac{\vec{p}\, ^2}{2M}
\end{eqnarray}
for the Galilei group and
\begin{eqnarray}
\label{kgenfunc}
     \vec{P} & = & \vec{p}, \quad \quad \, \vec{J} = \vec{q}\hspace{-0.1cm} \times \hspace{-0.1cm}\vec{p},    \nonumber \\
  \vec{K} & = & H\vec{q}, \quad H = \sqrt{\vec{p}\, ^2 + M^2}
\end{eqnarray}
for the Poincar\'e group. Checking that these are solutions of Eqs.~(\ref{bracrel})-(\ref{Kbracrel3}) is to be done as Problem 7.1.

Physical interpretation is established by identifying a three-dimensional vector function of the canonical coordinates and momenta that can represent the position of the object. We assume it  is changed by the canonical transformations generated by $H$, $\vec P$, $\vec J$, and $\vec G$ or $\vec K$ the way the position coordinate vector should be changed by the corresponding coordinate changes in the Galilei or Poincar\'e group. The only possibility is that $\vec q$ represents the position.\cite{me74} Then, in the nonrelativistic case, for the Galilei group, the velocity is
\begin{equation}
\label{gvel}
 \vec V = [\vec q, H] = \frac{\vec p}{M}
\end{equation}
so $\vec p = M\vec V$ is the momentum of an object with mass $M$ and velocity $\vec V$, the angular momentum $\vec{q}\hspace{-0.1cm} \times \hspace{-0.1cm}\vec{p}$ is $\vec J$, and
\begin{equation}
\label{gke}
 H =  \frac{1}{2}M\vec{V}\, ^2
\end{equation}
is the kinetic energy. In the relativistic case, for the Poincar\'e group, the velocity is
\begin{equation}
\label{rvel}
 \vec V = [\vec q, H] = \frac{\vec p}{\sqrt{\vec{p}\, ^2 + M^2}}
\end{equation}
so $\vec p = M\vec V/\sqrt{1 - \vec{V}\, ^2}$
is the relativistic momentum of an object with mass $M$ and velocity $\vec V$, the relativistic angular momentum $\vec{q}\hspace{-0.1cm} \times \hspace{-0.1cm}\vec{p}$ is $\vec J$, and
\begin{equation}
\label{rele}
 H =  \frac{M}{\sqrt{1 - \vec{V}\, ^2}}
\end{equation}
is the relativistic energy.

Conversely, if it is assumed that $\vec q$ represents the position of the object and is changed by the canonical transformations generated by $H$, $\vec P$, $\vec J$, and $\vec G$ or $\vec K$ the way the position coordinate vector should be changed by the corresponding coordinate changes in the Galilei or Poincar\'e group, then the generators $H$, $\vec P$, $\vec J$, and $\vec G$ or $\vec K$ can be put in the forms (\ref{ggenfunc}) or (\ref{kgenfunc}) by a canonical transformation that changes only the canonical momenta $\vec p$ and leaves the canonical coordinates $\vec q$ unchanged.\cite{me74} This is a gauge transformation.\cite{me74} 

For two objects, we use canonical coordinates and momenta $\vec q \, ^{(1)}$, $\vec q \, ^{(2)}$ and $\vec p \, ^{(1)}$, $\vec p \, ^{(2)}$. For the Galilei group, we can let
\begin{eqnarray}
\label{ggenfuncx2}
     \vec{P} & = & \vec p \, ^{(1)} + \vec p \, ^{(2)}, \quad \quad \quad \, \, \, \, \vec{J} = \vec q \, ^{(1)}\hspace{-0.1cm} \times \hspace{-0.1cm}\vec p \, ^{(1)} + \vec q \, ^{(2)}\hspace{-0.1cm} \times \hspace{-0.1cm}\vec p \, ^{(2)},    \nonumber \\[0.4cm]
  \vec{G} & = & m_1\vec q \, ^{(1)} + m_2\vec q \, ^{(2)}, \quad H = \frac{(\vec p \, ^{(1)})^2}{2m_1} + \frac{(\vec p \, ^{(2)})^2}{2m_2} + V
\end{eqnarray}
in which $V$ is the potential energy that describes the interaction between the two objects. The bracket relations for the Galilei group imply that $m_1 + m_2$ is $M$ and that $[V, \vec P]$, $[V, \vec G]$, and $[V, \vec J\, ]$ are zero. There are restrictions on $V$. To describe them, we use center-of-mass and relative coordinates and momenta 
\begin{eqnarray}
\label{cmrelqp}
     \vec{Q} & = & \frac{m_1}{m_1 + m_2}\vec q \, ^{(1)} + \frac{m_2}{m_1 + m_2}\vec q \, ^{(2)}, \quad  
\vec P \, ^{(tot)} = \vec p \, ^{(1)} + \vec p \, ^{(2)},    \nonumber \\[0.4cm]
  \vec{q} & = & \vec q \, ^{(1)} - \vec q \, ^{(2)}, \quad \vec p = \frac{m_2}{m_1 + m_2}\vec p \, ^{(1)} - \frac{m_1}{m_1 + m_2}\vec p \, ^{(2)}.
\end{eqnarray}
In terms of these, the generators (\ref{ggenfuncx2}) are
\begin{eqnarray}
\label{ggencmrel}
     \vec{P} & = & \vec P \, ^{(tot)}, \quad \, \, \, \,  \vec{J} = \vec Q\hspace{-0.1cm} \times \hspace{-0.1cm}\vec P + \vec q \hspace{-0.1cm} \times \hspace{-0.1cm}\vec p ,    \nonumber \\[0.4cm]
  \vec{G} & = & M\vec Q, \quad \quad H = \frac{\vec P \, ^2}{2M} + \frac{\vec p \, ^2}{2\mu } + V
\end{eqnarray}
with $\mu = m_1m_2/(m_1+m_2)$. We will write $\vec P$ for $\vec P \, ^{(tot)}$; it will be the same whether we are thinking about generators or canonical momenta. The $\vec Q$, $\vec P$, $\vec q$, $\vec p$ are canonical coordinates and momenta. Poisson brackets can be written with derivatives with respect to them. Thus the restrictions on $V$ from the bracket relations give
\begin{eqnarray}
\label{restrictv}
  \frac{\partial V}{\partial Q_j} & = & [V, P_j] = 0    \nonumber \\
  \frac{\partial V}{\partial P_j} & = & -\frac{1}{M}[V, G_j] = 0.
\end{eqnarray}
This means that $V$ can not depend on $\vec Q$ or $\vec P$; it can depend only on $\vec q$ and $\vec p$. Then $[V, \vec J\, ]$ is $[V, \vec q \hspace{-0.1cm} \times \hspace{-0.1cm}\vec p\, ]$. The bracket relations imply it is zero. It is to be shown in Problem 7.2 that this implies that $V$ is a function only of $\vec q \, ^2$, $\vec p \, ^2$ and $\vec q\cdot \vec p$. 

For the Poincar\'e group, describing interactions is not so simple. Generators can be written as sums of generators for single objects, describing objects that do not interact. If $H$ is changed to describe an interaction, either $\vec K$ or $\vec P$ must be changed too, because $H$ is $[K_k, P_k]$. Generators that describe interactions can be made to satisfy the bracket relations,\cite{Dirac49,Thomas1952,Thomas1953,Bakamjian1961,Foldy1961} but if it is also assumed that the position coordinates are changed by the canonical transformations the way they should be changed, corresponding to coordinate changes in the Poincar\'e group, then there are no interactions; the accelerations of all the objects are zero.\cite{me7,Currie1963,me9,Leutwyler1965} This shows that there are limits to the use of canonical transformations for a representation of the Poincar\'e group that contains a Hamiltonian description of the relativistic dynamics of different objects at the same time without fields. 
\vspace{0.6cm}

\noindent\textbf{Problem 7.1.} Show that the functions $H$, $\vec P$, $\vec J$, $\vec G$ given by Eqs.(\ref{ggenfunc}) satisfy the Poisson-bracket relations (\ref{bracrel}) - (\ref{Gbracrel3}) for the Galilei group and that the other Poisson brackets of these functions are zero. Show that the functions $H$, $\vec P$, $\vec J$, $\vec K$ given by Eqs.(\ref{kgenfunc}) satisfy the Poisson-bracket relations (\ref{bracrel}), (\ref{Kbracrel2}) and (\ref{Kbracrel3}) for the Poincar\'e group and that the other Poisson brackets of these functions are zero.
\vspace{0.6cm}

\noindent\textbf{Problem 7.2.} Consider changes of canonical coordinates and momenta $\vec q$ and $\vec p$ made by rotations around a fixed axis along the direction of a unit vector $\hat e$. To first order, for an infinitesimal value $\epsilon $ of the angle of rotation, 
\begin{eqnarray}
\label{qprot}
\vec q & \rightarrow  &  \vec q  +  \epsilon \, \hat e\hspace{-0.1cm} \times \hspace{-0.1cm} \vec q  \nonumber \\
\vec p & \rightarrow  &  \vec p  +  \epsilon \, \hat e\hspace{-0.1cm} \times \hspace{-0.1cm} \vec p
\end{eqnarray}
as in Eq.~(\ref{vecrot}). These changes of $\vec q$ and $\vec p$ form a one-parameter group of canonical transformations. The generator is $\hat e \hspace{-0.06cm}\cdot \hspace{-0.06cm} \vec q \hspace{-0.1cm} \times \hspace{-0.1cm}\vec p$; show this by showing that
\begin{eqnarray}
\label{qxpgen}
\hat e\hspace{-0.1cm} \times \hspace{-0.1cm} \vec q  & =  &  [\vec q, \,  \hat e \hspace{-0.06cm}\cdot \hspace{-0.06cm} \vec q \hspace{-0.1cm} \times \hspace{-0.1cm}\vec p ],
  \nonumber \\
\hat e\hspace{-0.1cm} \times \hspace{-0.1cm} \vec p  & =  &  [\vec p, \, \hat e \hspace{-0.06cm}\cdot \hspace{-0.06cm} \vec q \hspace{-0.1cm} \times \hspace{-0.1cm}\vec p ].
\end{eqnarray}
This implies that if $V$ is a function of $\vec q$ and $\vec p$ and $[V, \vec q \hspace{-0.1cm} \times \hspace{-0.1cm}\vec p\, ]$ is zero, then $V$ is not changed when $\vec q$ and $\vec p$ are rotated. This means that $V$ can only be a function of $\vec q \, ^2$, $\vec p \, ^2$ and $\vec q\cdot \vec p$. 
\vspace{0.6cm}

\noindent\textbf{Problem 7.3.} For one-dimensional space, the Galilei group has generators $H$, $P$, $G$, and bracket relations
\begin{equation}
\label{gbracrel1d}
 [P, H] = F, \quad [G, H] = P, \quad [G, P] = M
\end{equation}
with $F$ and $M$ constants. Neither $F$ nor $M$ can generally be eliminated by adding constants to the generators. Show this with an example. Specifically, consider a single object with canonical coordinates and momenta $q$ and $p$. Let $P$ be $p$, let $G$ be $Mq$, and find a function $H$ of $q$ and $p$ so that the bracket relations (\ref{gbracrel1d}) are satisfied with constants $F$ and $M$ that can not be removed. Show that then $p$ is $MV$ with $V$ the velocity $[q, H]$. Since $F$ is $[p, H]$, it is the time derivative of the momentum. It is a force. A constant force is allowed because Galilei transformations do not change accelerations. It is not allowed for three-dimensional space where there are also rotations. 
\vspace{0.6cm}

\noindent\textbf{Problem 7.4.} For one-dimensional space, the Poincar\'e group has generators $H$, $P$, $K$, and bracket relations
\begin{equation}
\label{kbracrel1d}
 [P, H] = F, \quad [K, H] = P, \quad [K, P] = H
\end{equation}
with $F$ a constant. This constant $F$ generally can not be eliminated by adding constants to the generators. Show this with an example. Specifically, consider a single object with canonical coordinates and momenta $q$ and $p$. Let $P$ be $p$. Find functions $H$ and $K$ of $q$, $p$ so that $p$ is $MV/\sqrt{1 - V^2}$, with $V$ the velocity $[q, H]$, and the bracket relations (\ref{kbracrel1d}) are satisfied with a constant $F$ that can not be removed. Since $F$ is $[p, H]$, it is the time derivative of the momentum. It is a force. A constant force is allowed because $dp'/dt'$ after a Lorentz transformation is the same as $dp/dt$ before. It is not allowed by Lorentz transformations for three-dimensional space. 
\vspace{0.6cm}

\noindent\textbf{Problem 7.5.} Consider a nonrelativistic particle with canonical coordinates and momenta $\vec{q}$ and $\vec{p}$ and generators of the Galilei group given by Eqs.(\ref{ggenfunc}). Suppose physical interpretation is established as described in the second and third paragraphs of this Section VII, so the velocity and energy are described by Eqs.(\ref{gvel}) and (\ref{gke}), the momentum $\vec p$ is $ M\vec V$, the angular momentum $\vec J$ is $\vec{q}\hspace{-0.1cm} \times \hspace{-0.1cm}\vec{p}$, and the mass is $M$. Show that $[G_j, P_k] = \delta_{jk}M$ implies that the momentum is changed by $-M\vec \beta $ when the space coordinates are changed by a Galilei transformation described by Eqs.(\ref{Gal}) to coordinates relative to an origin moving with velocity $\vec \beta $. Suppose all the particles are described this way in the decay of a particle of mass $m$ into two particles with masses $m_1$ and $m_2$. Show that conservation of momentum for the two coordinate systems with the origin of the second moving relative to the first implies that $m = m_1 + m_2$. That mass is not conserved in radioactive decays that were being observed before Einstein presented his relativity could have been seen as showing a need for revision of the relativity of Galileo and Newton. 

\appendix*

\section{From subgroups to generators}

Here we show that every one-parameter group of canonical transformations has a generator function that acts like a Hamiltonian. We use the canonical coordinates and momenta $q_n$ and $p_n$ together as variables we call $\zeta _j$. We do not need to distinguish the half of the variables $\zeta _j$ that are $q_n$ from the half that are $p_n$. The generator, which we call $H$, is to be a function of these variables. By letting $F$ be $\zeta _j$ in Eq.~(\ref{seriessolution}), we can see from the first-order term that the canonical transformations determine  
what the Poisson bracket $[\zeta _j, H]$ is to be. If there is an $H$, each $\partial H/\partial \zeta _k$ will be either $[\zeta _j, H]$ or $[-\zeta _j, H]$ for some $j$. We conclude that the canonical transformations determine what the $\partial H/\partial \zeta _k$ are to be. We do not assume there is a function $H$, so we can not write $[\zeta _j, H]$ and $\partial H/\partial \zeta _k$ for the things that are determined by the canonical transformations. We will write $[\zeta _j, H?]$ and $\partial H?/\partial \zeta _k$ for them until we have proved there is a function $H$.
The canonical transformations do not change the Poisson brackets $[\zeta _j,\zeta _k]$. This means that
\begin{eqnarray}
\label{zzbrac}
[\zeta _j,\zeta _k] & = & [\zeta _j + t[\zeta _j,H?] + ... \, ,\zeta _k + t[\zeta _k,H?] + ... \, ] \nonumber \\
& = & [\zeta _j,\zeta _k] + t[[\zeta _j,H?],\zeta _k] + t[\zeta _j,[\zeta _k,H?]] + ...
\end{eqnarray}
which implies that
\begin{equation}
\label{intcond}
\frac{\partial }{\partial \zeta _j}\frac{\partial H?}{\partial \zeta _k} = \frac{\partial }{\partial \zeta _k}\frac{\partial H?}{\partial \zeta _j}
\end{equation}
for all $j,k$. Let
\begin{equation}
\label{Hint}
H(\zeta ) = \int ^\zeta _{\zeta (0)} \sum _j\frac{\partial H?}{\partial \zeta '_j}d\zeta '_j.
\end{equation}
The $\zeta $ stands for the set of variables $\zeta _j$. The $\zeta (0)$ stands for a particular set of values of the $\zeta _j$, which mark a particular point in the ``phase'' space for which the $\zeta _j$ are coordinates. The integral is along a path in that space from the point $\zeta (0)$ to the point $\zeta $. We will show that the equations (\ref{intcond}) imply that the integral does not depend on the path. It defines $H$ as just a function of $\zeta $. It gives
\begin{equation}
\label{Hpar}
\frac{\partial H}{\partial \zeta _k} = \frac{\partial H?}{\partial \zeta _k}.
\end{equation}
Changing the point $\zeta (0)$ where the integral begins adds only a constant to $H$.

We can show that the integral is the same along two different paths because it is zero around the closed loop going forward on one path and back on the other. Let $\eta $ and $\lambda $ be variables that change along the two paths so that every point on the closed loop is marked by a different set of values of $\eta $ and $\lambda $. Along the loop, the $\zeta _j$ are functions of $\eta $ and $\lambda $ and they change only when $\eta $ or $\lambda $ changes. The integral (\ref{Hint}) around the closed loop is the integral over the area enclosed by the loop in the plane of $\eta $ and $\lambda $,
\begin{equation}
\label{loopint}
\oint (A_\eta d\eta  + A_\lambda d\lambda ) = \int  _{Area} (\frac{\partial A_\eta }{\partial \lambda } - \frac{\partial A_\lambda  }{\partial \eta  })d\eta \, d\lambda ,
\end{equation}
with
\begin{equation}
\label{Asubs}
A_\eta  = \sum _j\frac{\partial H?}{\partial \zeta _j}\frac{\partial \zeta _j}{\partial \eta }, \quad \quad
A_\lambda  = \sum _j\frac{\partial H?}{\partial \zeta _j}\frac{\partial \zeta _j}{\partial \lambda  },
\end{equation}
and
\begin{equation}
\label{cross}
\frac{\partial A_\eta }{\partial \lambda } - \frac{\partial A_\lambda  }{\partial \eta  }  = \sum _j\sum _k \Bigg( \frac{\partial }{\partial \zeta _j}\frac{\partial H?}{\partial \zeta _k} - \frac{\partial }{\partial \zeta _k}\frac{\partial H?}{\partial \zeta _j}\Bigg)\frac{\partial \zeta _k}{\partial \eta }\frac{\partial \zeta _j}{\partial \lambda },
\end{equation}
which is zero from Eq.~(\ref{intcond}).

\end{document}